\documentclass{sig-alternate}
\usepackage[T1]{fontenc}
\usepackage{lmodern}      
\usepackage[utf8]{inputenc}  
\usepackage{microtype}      
\usepackage{enumitem}
\usepackage{microtype}   
\usepackage{flushend}
\usepackage{ellipsis}   
\usepackage{mdframed}
\usepackage{hyperref}

\begin{document}
\toappear{}
%
%
\conferenceinfo{SSE'15}{September 1, 2015, Bergamo, Italy.}
\CopyrightYear{2015} 

\title{Understanding the Affect of Developers: Theoretical Background and Guidelines for Psychoempirical Software Engineering}

\numberofauthors{3}

\author{
\alignauthor
Daniel Graziotin\\
 \affaddr{Faculty of Computer Science, Free University of Bozen-Bolzano}\\
 \affaddr{Piazza Domenicani 3}\\
 \affaddr{Bolzano/Bozen, Italy}\\
 \email{daniel.graziotin@unibz.it}\\
\alignauthor 
Xiaofeng Wang\\
 \affaddr{Faculty of Computer Science, Free University of Bozen-Bolzano}\\
 \affaddr{Piazza Domenicani 3}\\
 \affaddr{Bolzano/Bozen, Italy}\\
 \email{xiaofeng.wang@unibz.it}\\
\alignauthor 
Pekka Abrahamsson\\
 \affaddr{Department of computer and information science}\\
 \affaddr{Norwegian University of Science and Technology}\\
 \affaddr{NO-7491 Trondheim, Norway}\\
 \email{pekkaa@ntnu.no}\\
}

\maketitle

\begin{abstract}
Affects---emotions and moods---have an impact on cognitive processing activities and the working 
performance of individuals. It has been established that software development tasks are undertaken 
through cognitive processing activities. Therefore, we have proposed to employ psychology theory and 
measurements in software engineering (SE) research. We have called it ``psychoempirical software engineering''. 
However, we found out that existing SE research has often fallen into misconceptions about the affect 
of developers, lacking in background theory and how to successfully employ psychological 
measurements in studies. The contribution of this paper is threefold. (1) It highlights the challenges
to conduct proper affect-related studies with psychology; (2) it provides a comprehensive literature
review in affect theory; and (3) it proposes guidelines for conducting psychoempirical software engineering.
\end{abstract}

\category{D.2.9}{Software Engineering}{Management}[Productivity, Programming Teams]
\category{H.1.2}{Models and Principles}{User/Machine Systems}[Human factors, Software psychology]
\category{J.4}{Social and behavioral Science}[Psychology]

\keywords{Affects, emotions, moods, human aspects in software development, psychology of programming, 
psychoempirical software engineering}

\section{Introduction}

The Nobel prize winner Daniel Kahneman has pointed out that it is unrealistic to limit our understanding of human behaviors solely through rational models \cite{Kahneman2003}. Yet, software engineering (SE) research has been known to be too much confined in the fallacy of rationality-above-everything paradigm \cite{Murgia2014}, to miss out the possibility to be a social discipline \cite{Sjoberg2008}, and to focus too much on domains of technical nature while neglecting the so-called \textit{soft aspects} or \textit{human-related} topics \cite{Lenberg2015}. But software development \textit{is} a very human activity. Software development happens in our minds first, then on artifacts \cite{Fischer1987}. It has been established that development is intellectual, and it is carried out through cognitive processing activities \cite{Feldt2010, Fischer1987, Khan2010}. Indeed, we are human beings, and, as such, we behave based on affect as we encounter the world through our emotions and moods \cite{Ciborra2002}. The affects pervade organizations by coloring the workers' thoughts, and they influence their behavior \cite{Brief2002}. Affects have a role in the relationships between workers, deadlines, work motivation, sense-making, and human-resource processes \cite{Barsade2007}. Although affects have been historically neglected in the studies of industrial and organizational psychology \cite{Muchinsky2000}, an interest in the role of affects on job outcomes has accelerated over the past fifteen years in psychology research \cite{Fisher2000c}. While research is still needed on the impact of affects on cognitive activities and work-related achievements in general, this link undeniably exists according to psychology research.

We have shown elsewhere \cite{Graziotin2014IEEESW} that practitioners are deeply interested in their affects while developing software, which causes them to engage in long and interesting discussions when reading related articles. Thus, it is important to understand the role of affects in software development processes. Even more, we share the view of Lenberg et al. \cite{Lenberg2015} that SE should also be studied from a behavioral perspective. We have, in fact, focused on these issues for some time now, by producing several articles on this avenue, i.e., \cite{Graziotin2013PROFES, Graziotin2013PROFESDOCSYM, Graziotin2015JSEP, Graziotin2014PEERJ, Graziotin2014IEEESW, Graziotin2015CHASE}. We have also proposed the term \emph{psychoempirical software engineering} \cite{Graziotin2014ISERN} to denote research in SE with proper theory and measurement from psychology. Our message was well-received by the community with some degree of agreements regarding terminology, e.g., \cite{Lenberg2014,Lenberg2015}. However, we show below that long is the road to properly address the human aspects of SE with psychology.

\paragraph{Problem: SE Lacks in Theoretical Background of Affects and Guidelines for Using Psychology}

Given the rising number of recent SE articles that deal with the affects of developers, e.g., \cite{Muller2015,Ford2015,Haaranen2015,Dewan2015}, we believe that it should be important for researchers to adopt a critical view of the phenomenon under study, and that they do not fall into the several misconceptions when dealing with the affect of developers \cite{Graziotin2015CHASE}. 

Yet, we understand that we have placed ourselves in a ``very confused and confusing field of study'' (\cite{Ortony1989}, p. 2). We experienced this confusion especially during our talks at ISERN 2014, where we chaired a workshop called psychoempirical SE \cite{Graziotin2014ISERN}, and during the CHASE 2015 workshop \cite{Begel2015}, where we presented some common misconceptions and measurements of the affect of software developers  \cite{Graziotin2015CHASE}. Such misconceptions include confusing affect and the related constructs of emotions and moods with motivation or job satisfaction, which has happened even in articles already dealing with misconceptions of motivation with respect to job satisfaction, e.g., \cite{Franca2014}, although affects were not the focus of the study in this case. 

Other issues lie in missing out the opportunity of using validated measurement instruments for affect. An example is the use of the niko-niko calendar for assessing the mood of a software development team, e.g. \cite{Sato2006}, or the so-called happiness index, e.g., \cite{Medinilla2014}. Another example of the missed opportunity is when a single truth is assumed in the writing of articles, like in a CACM positional article claiming that ``psychologist recognize eight basic emotions, with each positive balanced by a negative'', e.g. ``love-hate'' (\cite{Denning2012b}, p. 34), or in a proper empirical study where it has been claimed that ``there are six basic emotions or universal emotions: anger, happiness, fear, [..]'' (\cite{Colomo-Palacios2013}, p. 1079). We will show below that it is not true that a unique dominant, accepted theory exists for affect, emotions, and moods. Researchers should recognize this issue when employing such delicate concepts for conducting research. 

\paragraph{Proposal: Theoretical Background of Affects and Guidelines for Psychoempirical SE}

While it would be preposterously arrogant on our side to claim the all-encompassing knowledge of the topic, we would like to share what we have learned so far in our journey to understanding software developers through their affect. This article builds upon our experience, the feedback collected at our talks and peer review processes, and the previously conducted research, to build some theoretical background for understanding the affect of software developers. We draw from research in psychology in the last decades, and offer a comprehensive review of the theory of affect (section \ref{sec:affect_emotions_moods_theoretical_background}) and, as a follow-up of our ISERN 2014 workshop \cite{Graziotin2014ISERN}, we propose our guidelines for psychoempirical SE (Section \ref{sec:guidelines_for_psychoempirical_software_engineering}) for conducting studies in SE with psychological theory and measurement.

\section{Affect, Emotions, Moods: Theoretical Background}
\label{sec:affect_emotions_moods_theoretical_background}

The fields of psychology have failed to agree on the definitions of affects and the related terms such as
emotions, moods, and feelings \cite{Ortony1989, Russell2003}. Yet, it is desirable that we provide
a starting set of definitions, which we will however criticize.

Let us start by stating that the term \emph{affect} (or
affective state) has been defined as ``any type of emotional state
{[}\ldots{}{]} often used in situations where emotions dominate the
person's awareness'' \cite{VandenBos2013}. This definition is problematic as
it contains the term \emph{emotion}, which has not yet been defined, and
it does not help in defining the (now apparently) super-construct \emph{affects}. 
Indeed, the term \emph{affects} is often associated in the
literature with \emph{emotions} and \emph{moods}. We now are left with three terms, which look remarkably similar to each other.

Plutchik \cite{Plutchik1980} has defined \emph{emotions} as the states
of mind that are raised by external stimuli and are directed toward the
stimulus in the environment by which they are raised. However, 
Kleinginna et al. \cite{Kleinginna1981} reported one year later that more than 90 definitions have been produced for this term, and no consensus in the literature has been reached. The term has been taken for granted and
often defined with references to a list, e.g. anger, fear, joy, surprise 
\cite{Cabanac2002}. To worsen this, \emph{emotion} as a term is not universally
employed, as it is a word that does not exist in all languages and
cultures \cite{Russell1991}.

\emph{Moods} have been defined as emotional states in which the
individual feels good or bad, and either likes or dislikes what is
happening around him/her \cite{Parkinson1996}. Yet again, a definition
of one construct contains another construct of our interest.

\paragraph{How Do Emotions and Moods Differentiate, Then?} While for some researchers certain moods are emotions and vice versa \cite{DeLancey2006}, it has been suggested that a distinction is not necessary for studying
cognitive responses that are not strictly connected to the origin of the
mood or emotion \cite{Weiss1996}. Distinctions between emotions and moods are clouded, because both may feel very much the same from the perspective of an individual experiencing either \cite{Beedie2005} and are now a part of common sense \cite{Russell2003}. They are embedded in psychologists' questions and, as a consequence, answers. Reisenzein \cite{Reisenzein2007} argued that ``\emph{the consensual definition of emotion is not a precondition but the result of scientific research; and even then, it remains a revisable empirical
hypothesis}'' (p. 2). So, affects, emotions, and moods are an emergent construction rather than a 
latent entity \cite{Clore2008, Minsky2008}. 

We have adopted the same stance of several researchers in the various
fields \cite{Schwarz1983, Schwarz1990, Wegge2006b, DeDreu2011} and employed the noun \emph{affects} (affective states) as an umbrella term for emotions and moods. We will show that, according to a recent unifying theory, this strategy does make sense.

\subsection{The Major Frameworks for Affect Theories}
\label{ssec:2literaturereview:the_two_major_frameworks_for_affect_theories}

According to Huang \cite{Huang2001}, four major theories exist for emotions
(moods, affects) in psychology. However, we see that these four theories
and all the other we could review fall into two competing frameworks.

\paragraph{The Discrete Framework}

One framework, namely the discrete approach, collects a set of basic
affective states that can be distinguished uniquely \cite{Plutchik1980}, and 
that possess high cross-cultural agreement when
evaluated by people in literate and preliterate cultures \cite{Ekman1971}.

The Differential Emotions Theory \cite{Izard1977} states that the human
motivation system is based on ten fundamental emotions (interest, joy,
surprise, distress, anger, disgust, contempt, fear, shame, and guilt).
These fundamental emotions function for the survival of human beings,
possess an own neural network in the brain, and an own behavioral
response. Finally, these emotions interact with each other
simultaneously.

Ekman \cite{Ekman1971} proposed a set of basic affects, which include anger,
happiness, surprise, disgust, sadness, and fear. However, the list has
received critique, leading to an extended version of eleven
elements \cite{Ekman1992}. They include amusement, embarrassment, relief,
and shame.

In the Circular Model of Emotion \cite{Plutchik1980}, a 
structure describing the interrelations among emotions has been
proposed. Eight primary, bipolar affective states were presented as
coupled pairs: joy versus sadness, anger versus fear, trust versus
disgust, and surprise versus anticipation. These eight basic emotions
vary in intensity and can be combined with each other, to form secondary
emotions. For example, joy has been set as the midpoint between serenity
and ecstasy, whereas sadness has been shown to be the midpoint between
pensiveness and grief. Emotions can vary in intensity and persistence
(to form moods, for example). Emotions, under this theory, serve an
adaptive role in dealing with survival issues. 

Developing a minimal list of basic affective states appears to be difficult with the discrete approach.
Subsequent studies have come to the point where more than 100 basic emotions have been proposed \cite{Shaver1987}.

\paragraph{The Dimensional Framework}

\begin{figure}[t]
\centering
\includegraphics[width=\columnwidth]{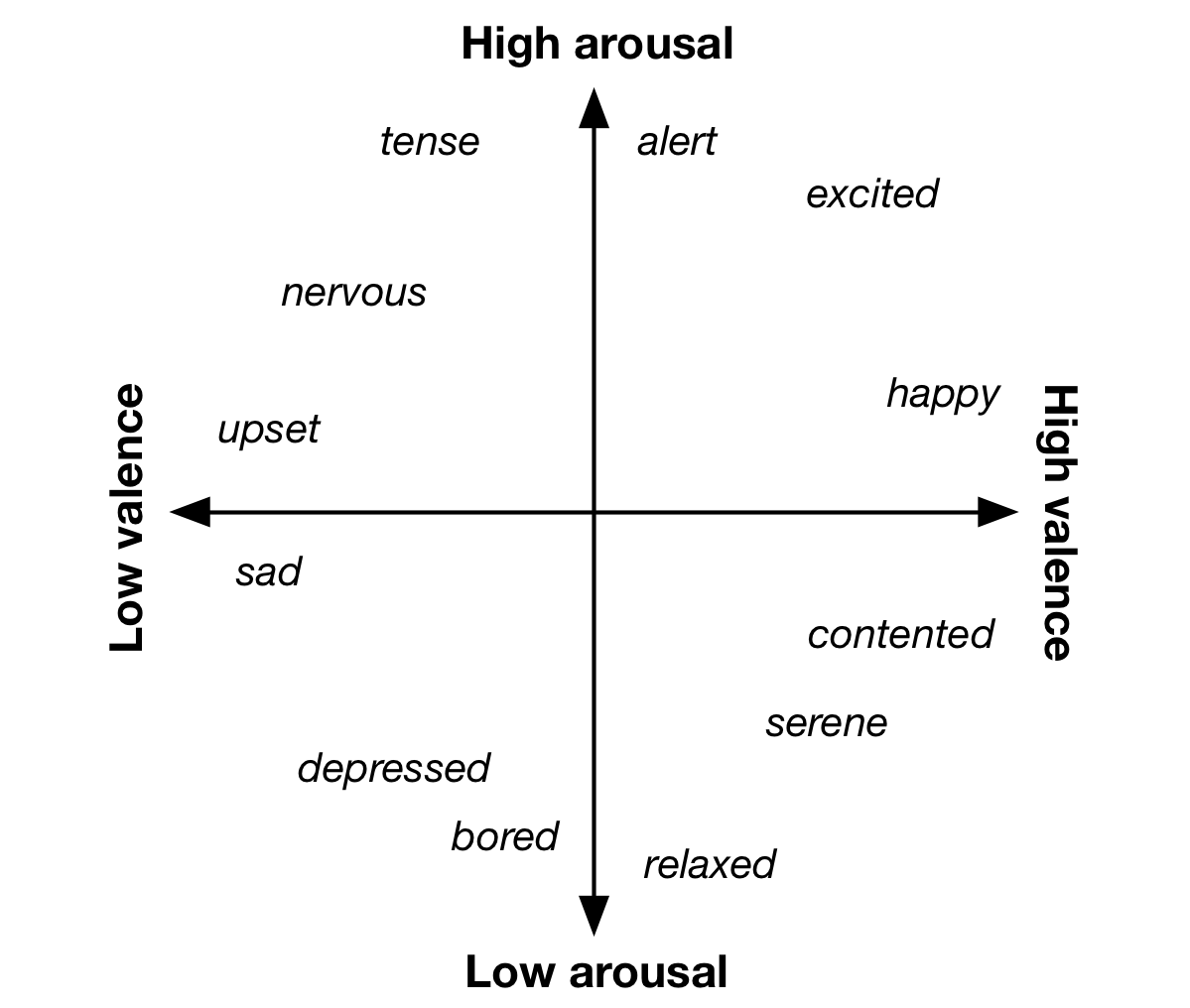} 
\caption{A PAD model of valence and arousal, and examples of related discrete affects.}\label{fig:pad}
\end{figure}

The dimensional framework groups affects in major dimensions that
allow a clear distinction among them \cite{Russell1980, Lane1999}. In the PAD models, three dimensions of Pleasure-displeasure, Arousal-nonarousal, and Dominance-submissiveness \cite{Russell1977,Russell1980, Mehrabian1996} characterize the emotional states of humans. \emph{Valence} (or pleasure) is the attractiveness (or adverseness) of an event, object, or situation \cite{Lewin1935} \cite{Lang1993}. The term refers to the ``direction of a behavioral activation associated toward (appetitive motivation) or away (aversive motivation) from a stimulus'' \cite{Lane1999}. \emph{Arousal} represents the
intensity of emotional activation \cite{Lane1999}. It is the sensation
of being mentally awake and reactive to stimuli, i.e. vigor and energy
or fatigue and tiredness \cite{Zajenkowski2012}. \emph{Dominance} (or
control, over-learning) represents a change in the sensation of the
control of a situation \cite{Lang1994}. It is the sensation by
which an individual's skills are perceived to be higher than the
challenge level for a task \cite{Csikszentmihalyi1997}. Figure \ref{fig:pad} provides a representation of a PAD model of valence and arousal, and examples of related discrete affects with an indication to where they might correspond on the axes.

Emotional states under the PAD models include moods, feelings, and any other
feeling-related concepts. The  dimensions are usually bipolar, indicating
that the presence of pleasure excludes the possibility of displeasure.
Some variations of these models have been proposed using different
notations but without changing the core meaning \cite{Russell2003}, some of
which omit the dominance dimension \cite{Lane1999}.

In the Positive and Negative Affect Schedule (PANAS) \cite{Watson1988, Watson1988b}, the positive and negative
affects are considered as the two primary emotional dimensions. However,
these two dimensions are the result of the self-evaluation of a number
of words and phrases that describe different feelings and emotions. That
is: discrete emotions are rated but two dimensions are evaluated. This
theory is designed to present a mood scale. Finally, positive and
negative affects are mutually independent. In Figure \ref{fig:pad}, the positive (negative) dimension would comprise of positive (negative) valence, or positive arousal, or both according to the different theories.

We note here that several other theories exist, although they are less prominent. One example is the cube of emotion \cite{Lovheim2012}, which is a dimensional theory of affect that expresses affect in terms of combinations of dopamine, adrenaline, and serotonin, which intersect in eight basic (but extreme) affects, e.g. distress, interest, joy.

\paragraph{The Unifying Theory}

A prominent unifying theory exists as well. Russell and Barrett \cite{Russell1999, Russell2003, Russell2009} have proposed the concept of \emph{core affect} to unify the theories of emotions and moods in psychology. Core affect is ``a pre-conceptual primitive process, a neurophysiological state, accessible to consciousness as a simple non-reflective feeling that is an integral blend of hedonic (valence) and arousal values: feeling good or bad, feeling lethargic or
energized'' \cite{Russell2003} (p. 147). The state is accessible at a consciousness level as the simplest raw feelings, which is distinct in moods and emotions. A feeling is an assessment of one's current condition. Therefore, an affect is a very raw concept, upon which the more complex of mood and emotion is built upon. \emph{Pride} can be thought of as feeling
good about oneself. The ``feeling good'' is core affect and the ``about
oneself'' is an additional (cognitive) component. 

Changes in affects result from a combination of happenings, such as stressful events on the job. Sometimes the cause of the change is obvious. However, sometimes one can undergo a change in core affect without understanding the reasons. The individuals possess a limited ability to track this complex causality connection. Instead, a person makes attributions and interpretations of core affect.

Affect can be felt in relation to no obvious stimulus---in a free-floating form---as \emph{moods} are perceived. Indeed, mood is defined as a prolonged core affect without an object, i.e. an unattributed affect. 

In the core affect theory, \emph{emotions} are episodes instead of simple objects. An emotion is a complex set of interrelated sub-events about a specific object.

The core affect theory is interesting because it unifies the previous theories, and it maintains compatibility with the majority of the existing measurement instruments, regardless of them being about moods or emotions. Although we do not neglect moods and emotions \emph{per se}, when adopting the core affect theory we chose to understand the states of minds of software developers at the \emph{affective} level only, which is the foundation of moods and emotions. 

\paragraph*{Core Affect Is our Current Suggestion to Frame SE Research on Affect} However, a researcher should select the affective framework and theory that better suits the research objective and the level of details that are desirable. We provide more details in the next section. What is important is that researchers are aware of an absence of an absolute truth and of the many existing alternatives, and that they justify their choice.

\section{Guidelines for Psychoempirical Software Engineering}
\label{sec:guidelines_for_psychoempirical_software_engineering}

A much requested feature in our previous discussions at recent academic venues such as ISERN 2014, CHASE 2015 \cite{Begel2015}, and ICSE 2015 had been \emph{How should one conduct research with psychological measurements?} By making sense of the hundreds of articles we reviewed on psychology and organizational behavior, we came up with a simple series of steps, listed below.

\paragraph{Defining a Research Objective}
As with any research activity, it is important to understand what we want to do in a study. Suppose two different, yet common scenarios with the affects of developers. They have been adapted from two of our previous studies \cite{Graziotin2014PEERJ,Graziotin2015JSEP}.

\begin{description}
\item[Scenario A] Assessing how happy developers are generally.
\item[Scenario B] Assessing over a time frame the emotional reaction of a stimulus (e.g., employing a software tool) on developers.
\end{description}

Both of them require a deep understanding of the topic under study.

\paragraph{Theoretically Framing the Research}

\emph{Scenario A}---From a comprehensive literature review, we would understand that we can call happy those developers who are in a strongly positive mood, or those who frequently have positive and meaningful experiences (see \cite{Graziotin2015CHASE} for more), thus having a positive affect balance. We decide to focus on dimensions of affects, e.g. with the Positive and Negative Affect Schedule (PANAS) \cite{Watson1988, Watson1988b}, which still lets us evaluate discrete affects before the aggregated scores.

\emph{Scenario B}---Suppose that, instead of asking a developer what emotions she is feeling when using a tool, we are interested in knowing how she feels in terms of more aggregated dimensions like pleasure, energy, and dominance. We focus then on the dimensional theory of affects like the one in the PAD models \cite{Russell1977,Russell1980, Mehrabian1996}.

\paragraph{Selecting a Validated Measurement Instrument}

\emph{Scenario A}---The PANAS dimensional model recommend employing the PANAS \cite{Watson1988, Watson1988b} 
measurement instrument which is one of the most notable measurement instruments for affective states. However, a deeper look at the literature shows that there are several shortcomings that have been criticized for this instrument. The PANAS reportedly omits core emotions such as \emph{bad} and \emph{joy} while including items that are not considered emotions, like \emph{strong}, \emph{alert}, and \emph{determined} \cite{Diener2009, Li2013}. Another limitation
has been reported in its non-consideration of the differences in
desirability of emotions and feelings in various cultures \cite{Tsai2006, 
Li2013}. Furthermore, a considerable redundancy has been found in PANAS items \cite{Crawford2004, Thompson2007, Li2013}. PANAS has also  been reported to capture only high-arousal feelings in general \cite{Diener2009}.

Recent, modern scales have been proposed to reduce the number of the
PANAS scale items and to overcome some of its shortcomings. Diener \cite{Diener2009} developed the Scale of Positive and Negative Experience
(SPANE). SPANE assesses a broad range of pleasant and unpleasant
emotions by asking the participants to report them in terms of their
frequency during the last four weeks. It is a 12-items scale, divided
into two sub-scales. Six items assess positive affective states and form
the SPANE-P scale. The other six assess negative affective states and
form the SPANE-N scale. The answers to the items are given on a
five-point scale ranging from 1 (\emph{very rarely or never}) to 5
(\emph{very often or always}). For example, a score of five for the
\emph{joyful} item means that the respondent experienced this affective
state \emph{very often} or \emph{always} during the last four weeks. The
SPANE-P and SPANE-N scores are the sum of the scores given to their
respective six items. Therefore, they range from 6 to 30. The two scores
can be further combined by subtracting SPANE-N from SPANE-P, resulting
in the Affect Balance Score (SPANE-B). SPANE-B is an indicator of the
pleasant and unpleasant affective states caused by how often positive
and negative affective states have been felt by the participant. SPANE-B
ranges from -24 (\emph{completely negative}) to +24 (\emph{completely
positive}). The SPANE measurement instrument has been reported to be
capable of measuring positive and negative affective states regardless
of their sources, arousal level or cultural context, and it captures
feelings from the emotion circumplex \cite{Diener2009, Li2013}.
The timespan of four weeks was chosen in SPANE in order to provide a
balance between the sampling adequacy of feelings and the accuracy of
memory \cite{Li2013}, and to decrease the ambiguity of people's
understanding of the scale itself \cite{Diener2009}. 

\begin{figure}[t]
\centering
\includegraphics[width=\columnwidth]{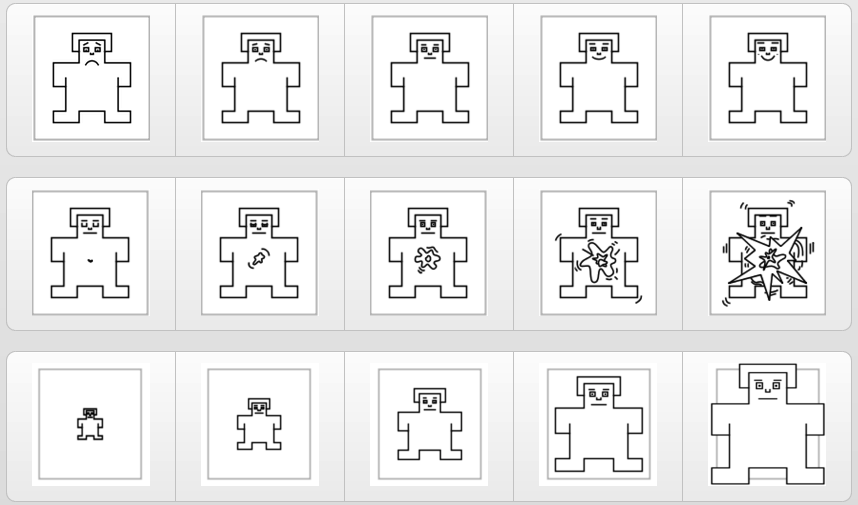} 
\caption{The Self-Assessment Manikin.}\label{fig:sam}
\end{figure}

\emph{Scenario B}---The PAD dimensional models have been implemented in several measurement instruments. 
One of the most notable instruments is the Affect Grid \cite{Russell1989}, which is a grid generated 
by intersecting the axes of valence and arousal accompanied by four discrete affects, 
i.e. depression-relaxation and stress-excitement, to guide the participant in pointing where the emotional 
reaction is located. The affect grid has been employed in SE research, e.g. in \cite{Colomo-Palacios2011}. 
Yet, the grid was shown to have only moderate validity \cite{Killgore1998}, 
thus other measurement instruments would be more desirable. 
Thus comes the Self-Assessment Manikin (SAM, \cite{Lang1994, Lang1999}). SAM is a pictorial,
i.e. non-verbal, assessment method. SAM measures valence, arousal, and
dominance associated with a person's affective reaction to an object (or
a stimulus) \cite{Lang1994}. As a picture is worth a thousand words, we reproduce SAM in figure \ref{fig:sam}. 
The figures of the first row range from a frown to a smile, representing the valence dimension. 
The second row depicts a figure showing a serene, peaceful, or passionless face to an explosive, anxious, or excited face. It represents the arousal dimension. The third row ranges from a very little, insignificant figure to a ubiquitous, pervasive figure. It represents 
the dominance affective dimension. As reported in \cite{Morris2002}, SAM has the advantage of eliminating
 the cognitive processes associated with verbal measures but it is still very quick and simple to use.

\paragraph{Considering Psychometric Properties}

As we noted in a previous paper \cite{Graziotin2015CHASE}, a selected measurement instrument has to possess acceptable validity and reliability properties, which are provided in psychometric studies of the measurement instrument. Psychometrics is a term, which has been misused in SE including ourselves. It is a subfield of psychology that focuses on the theory and techniques of psychological measurements. Psychometric studies deal with the design, development and especially the validation of psychological measures.

A modification to an existing measurement instrument (e.g., adding, deleting, or rewording items) often requires a new psychometric study because the reliability of a measurement instrument can be compromised. Therefore, it is not advisable to modify validated psychological measurements or models as it happened in \cite{Colomo-Palacios2013}.

\emph{Scenario A}---The SPANE has been validated to converge with other affective states measurement instruments, including PANAS \cite{Diener2009}. 
The scale provided good psychometric properties in the introductory research \cite{Diener2009} and 
in numerous follow-ups, with up to twenty-one thousand participants in a single study \cite{Silva2011, Dogan2013, Li2013}. 
Additionally, the scale proved consistency across full-time workers and students \cite{Silva2011}.

\emph{Scenario B}---The SAM has been under scholarly scrutiny, as well. The original article describing SAM already reports good psychometric properties \cite{Lang1994}. A very high correlation was found between the SAM items and those of other verbal-based measurement instruments \cite{Morris1993, Morris1995}, including high reliability across age \cite{Backs2007}. Therefore, SAM is one of the most reliable measurement instruments for affective reactions \cite{Morris2002}.

\paragraph{Administering the Measurement Instrument Correctly}

The psychometric properties of a measurement instrument in psychology are also calculated by administering the instrument in the same way in each study. This is because the instructions might influence the participants' responses. For this reason, any good measurement instrument is always accompanied with the instructions for the participants. We encourage administering a measurement instrument as it is reported in the accompanying instructions, and to further share the instructions with participants. Furthermore, the gained transparency ensures a higher reproducibility of the studies.

We strongly encourage the authors of SE studies to report the participants' instructions when publishing an article, preferably in an archived format. \footnote{
    For the participants' instructions in \cite{Graziotin2014PEERJ}, see \url{https://dx.doi.org/10.7717/peerj.289/supp-1}.
    For the participants' instructions in \cite{Graziotin2013PROFES,Graziotin2015JSEP}, see \url{http://dx.doi.org/10.6084/m9.figshare.796393}
}

\emph{Scenario A}---The SPANE instructions for participants are clearly stated in the original paper \cite{Diener2009} 
and in the instrument itself, which is freely available. \footnote{ \url{http://internal.psychology.illinois.edu/~ediener/SPANE.html}}

\emph{Scenario B}---The SAM instructions for participants are exhaustively reported in the accompanying technical report \cite{Lang1999}.

\paragraph{Performing Strong Analyses}

We encourage the authors in SE to spend some time to understand whether such complex and delicate constructs require accurate analyses.

\emph{Scenario A}---The SPANE scores can be considered as ordinal values or as discrete pinpoints of a continuous scale. Regression analyses on the aggregated SPANE-P, SPANE-N, and SPANE-B scores are possible given that the assumptions for linear regression are met. Otherwise, especially when groups have to be compared, the usual assumptions for employing the \emph{t-test} or non-parametric tests should be taken into account. It is also important to report an effect size measure such as the Cohen's $d$.

\emph{Scenario B}---Repeated measures within-subject that need a between subject comparison pose several issues. First, there is not a stable and shared metric for assessing the affects across persons. For example, a score of one in valence for a person may be equal to a score of three for another person. However, a participant scoring two for valence at time \emph{t} and five at time \emph{t+x} unquestionably indicates that the participant's valence
increased. As stated by Hektner \cite{Hektner2007}, ``\emph{it is sensible to assume that there is a reasonable stable metric within persons}'' (p. 10). In order to have comparable measurements, the raw scores of each participant are typically transformed into \emph{z-scores} (also known as standard scores). A z-score transformation is such that a participant's mean score for a variable is
zero, and scores for the same variable that lie one standard deviation
above or below the mean have the value equivalent to their deviation.
One observation is translated to how many standard deviations the
observation itself is above or below the mean of the individual's
observations. Therefore, the participants' measurements become
dimensionless and comparable with each other, because the z-scores
indicate how much the values are spread \cite{Larson1983, Hektner2007}.

Second, the repeated measurements often present dependencies of data at the participants'
level and the time level grouped by the participant. The analysis of
variance (ANOVA) family provides rANOVA as a variant for repeated
measurements. However, rANOVA and general ANOVA procedures are
discouraged \cite{Gueorguieva2004} in favor of mixed-effects
models, which are robust and specifically designed for repeated,
within-participant longitudinal data \cite{Laird1982, Gueorguieva2004, 
Baayen2008}. A linear mixed-effects model is a linear model that contains both fixed effects and random effects \cite{Robinson1991}. The estimation of the significance of the effects for mixed models is an open debate \cite{Bates2006, RCommunity2006}. We encourage the reader to follow our reasoning in \cite{Graziotin2015JSEP} for a deeper discussion.

\section{Conclusion}
Affects---emotions and moods---are beginning to be comprehensively studied in SE, and other psychological constructs are being incorporated in related research. However, there is a risk of underusing and misusing the theory and the measurement instruments from psychology, and falling into the many misconceptions tied to such intriguing and complex research topics.

For this reason, we have proposed the term \emph{psychoempirical software engineering} to refer to the
research in SE with psychology theory and measurement. This paper described the challenge to conduct 
proper affect-related studies with psychology, provided a comprehensive literature 
review in affect theory, and proposed guidelines for conducting psychoempirical software engineering.

With this article, we hope to raise much needed awareness for better use of psychology in SE studies and to begin a sane discussion with our peers towards a more standard and sound way of conducting studies on the human and social aspects of SE.

\section{Acknowledgments}
We are grateful for the comments received by all at the ISERN 2014 and CHASE 2015 workshops. We are also grateful to every member of the scientific community who spent time in reviewing our articles on the affect of developers, including the present one, which brought us to research in deep about this fascinating topic that deserves to be properly researched in SE.

\bibliographystyle{abbrv}
\bibliography{references}

%
%

\newpage

\onecolumn

\noindent
\fbox{
    \parbox{\textwidth}{
        D. Graziotin, X. Wang, and P. Abrahamsson. Understanding the affect of developers: theoretical background and guidelines for psychoempirical software engineering. In Proceedings of the 7th International Workshop on Social Software Engineering - SSE 2015, pages 25–32, New York, New York, USA, 2015. ACM Press. DOI: 10.1145/2804381.2804386.
    }
}

\vspace{10 mm}

\textcopyright the authors, 2015. This is the author's version of the work. It is posted here for your personal use. Not for redistribution. The definitive Version of Record was published in Proceedings of the 7th International Workshop on Social Software Engineering - SSE 2015, pages 25–32, New York, New York, USA, 2015. ACM Press. DOI: 10.1145/2804381.2804386.

\end{document}